
%
\catcode`@=11
%
\hyphenchar\tentt = -1
\def\tenpoint{%
   \textfont0=\tenrm \scriptfont0=\sevenrm
      \scriptscriptfont0=\fiverm
   \def\rm{\fam\z@ \tenrm}
   \textfont1=\teni \scriptfont1=\seveni
      \scriptscriptfont1=\fivei
   \def\mit{\fam\@ne} \def\oldstyle{\fam\@ne\teni}
   \textfont2=\tensy \scriptfont2=\sevensy
      \scriptscriptfont2=\fivesy
   \def\cal{\fam\tw@}
   \textfont3=\tenex \scriptfont3=\tenex
      \scriptscriptfont3=\tenex
   \textfont\bffam=\tenbf \scriptfont\bffam=\sevenbf
      \scriptscriptfont\bffam=\fivebf
   \def\bf{\fam\bffam\tenbf}
   \textfont\itfam=\tenit \def\it{\fam\itfam\tenit}
   \textfont\slfam=\tensl \def\sl{\fam\slfam\tensl}
   \textfont\ttfam=\tentt \def\tt{\fam\ttfam\tentt}
\def\big##1{{\hbox{$\left##1\vbox to 8.5pt{}\right.\n@space$}}}%
\def\Big##1{{\hbox{$\left##1\vbox to 11.5pt{}\right.\n@space$}}}%
\def\bigg##1{{\hbox{$\left##1\vbox to 14.5pt{}\right.\n@space$}}}%
\def\Bigg##1{{\hbox{$\left##1\vbox to 17.5pt{}\right.\n@space$}}}%
%
   \setbox\strutbox=\hbox{\vrule height 8.5pt depth 3.5pt width0pt}%
   \normalbaselineskip=12pt
   \normalbaselines\rm
}%
%
%
   \font\twelverm=cmr12
   \font\ninerm=cmr9
   \font\twelvei=cmmi12
   \font\ninei=cmmi9
   \font\twelvesy=cmsy10 scaled 1200
   \font\ninesy=cmsy9
   \font\twelveex=cmex10 scaled 1200
   \font\twelvebf=cmbx12
   \font\ninebf=cmbx9
   \font\twelvesl=cmsl12
   \font\twelveit=cmti12
   \font\twelvett=cmtt12
   \skewchar\twelvei='177
   \skewchar\ninei='177
   \skewchar\twelvesy='60
   \skewchar\ninesy='60
   \hyphenchar\twelvett=-1
\def\twelvepoint{%
   \textfont0=\twelverm \scriptfont0=\ninerm
      \scriptscriptfont0=\sevenrm
   \def\rm{\fam\z@ \twelverm}
   \textfont1=\twelvei \scriptfont1=\ninei
      \scriptscriptfont1=\seveni
   \def\mit{\fam\@ne} \def\oldstyle{\fam\@ne \twelvei}
   \textfont2=\twelvesy \scriptfont2=\ninesy
      \scriptscriptfont2=\sevensy
   \def\cal{\fam\tw@}
   \textfont3=\twelveex \scriptfont3=\twelveex
      \scriptscriptfont3=\twelveex
   \textfont\bffam=\twelvebf \scriptfont\bffam=\ninebf
      \scriptscriptfont\bffam=\sevenbf
   \def\bf{\fam\bffam\twelvebf}
   \textfont\itfam=\twelveit \def\it{\fam\itfam\twelveit}
   \textfont\slfam=\twelvesl \def\sl{\fam\slfam\twelvesl}
   \textfont\ttfam=\twelvett \def\tt{\fam\ttfam\twelvett}
\def\big##1{{\hbox{$\left##1\vbox to 10.2pt{}\right.\n@space$}}}%
\def\Big##1{{\hbox{$\left##1\vbox to 13.8pt{}\right.\n@space$}}}%
\def\bigg##1{{\hbox{$\left##1\vbox to 17.4pt{}\right.\n@space$}}}%
\def\Bigg##1{{\hbox{$\left##1\vbox to 21pt{}\right.\n@space$}}}%
%
   \setbox\strutbox=\hbox{\vrule height 10.2pt depth 4.2pt width0pt}%
   \normalbaselineskip=14.4pt
   \normalbaselines\rm
}%
%
%
\def \definefourteen {%
   \font \fourteenrm =cmr12 scaled 1200
   \font \elevenrm =cmr9 scaled 1200
   \font \fourteeni =cmmi12 scaled 1200
   \font \eleveni =cmmi9 scaled 1200
   \font \fourteensy =cmsy10 scaled 1400
   \font \elevensy =cmsy9 scaled 1200
   \font \fourteenex =cmex10 scaled 1400
   \font \fourteenbf =cmbx12 scaled 1200
   \font \elevenbf =cmbx9 scaled 1200
   \font \fourteensl =cmsl12 scaled 1200
   \font \fourteenit =cmti12 scaled 1200
   \font \fourteentt =cmtt12 scaled 1200
   \skewchar \fourteeni ='177
   \skewchar \eleveni ='177
   \skewchar \fourteensy ='60
   \skewchar \elevensy ='60
   \hyphenchar \fourteentt =-1
}%
\def \fourteenpoint {%
   \definefourteen
   \textfont0 =\fourteenrm \scriptfont0 =\elevenrm
      \scriptscriptfont0 =\elevenrm
   \def \rm {\fam0 \fourteenrm }%
   \textfont1 =\fourteeni \scriptfont1 =\eleveni
      \scriptscriptfont1 =\eleveni
   \def\mit{\fam\@ne} \def\oldstyle{\fam\@ne\fourteeni}
   \textfont2 =\fourteensy \scriptfont2 =\elevensy
      \scriptscriptfont2 =\elevensy
   \def\cal{\fam\tw@}
   \textfont3 =\fourteenex \scriptfont3 =\fourteenex
      \scriptscriptfont3 =\fourteenex
   \textfont \bffam =\fourteenbf \scriptfont \bffam =\elevenbf
      \scriptscriptfont \bffam =\elevenbf
   \def \bf {\fam \bffam \fourteenbf }
   \textfont \itfam =\fourteenit \def \it {\fam \itfam \fourteenit }
   \textfont \slfam =\fourteensl \def \sl {\fam \slfam \fourteensl }
   \textfont \ttfam =\fourteentt \def \tt {\fam \ttfam \fourteentt }
\def \big ##1 {{\hbox {$\left ##1\vbox to 11.9pt {}\right .\n@space $}}}%
\def \Big ##1 {{\hbox {$\left ##1\vbox to 16.1pt {}\right .\n@space $}}}%
\def \bigg ##1 {{\hbox {$\left ##1\vbox to 20.3pt {}\right .\n@space $}}}%
\def \Bigg ##1 {{\hbox {$\left ##1\vbox to 24.5pt {}\right .\n@space $}}}%
   \setbox \strutbox =\hbox {\vrule height 11.9pt depth 4.9pt width 0pt }%
   \normalbaselineskip =16.8pt
   \normalbaselines \rm
 }%
\catcode`@ =12 
\hfuzz=30 pt

\def\Re{\,{\rm Re}\,}
\def\Im{\,{\rm Im}\,}

\def\e{\, {\rm e}}
\def\IR{{\rm I}\!{\rm R}}
\def\Gamit{{\Gamma^{2n}}}
\def\MG2{\Gamma^2}
\def\zbar{{\overline{z}}}
\def\tbar{{\overline{\tau}}}
\def\inbar{\,\vrule height1.5ex width.4pt depth0pt}
\def\IC{\relax\hbox{$\inbar\kern-.3em{\rm C}$}}
\def\IZ{\relax\ifmmode\mathchoice{\hbox{Z\kern-.4em Z}}{\hbox{Z\kern-.4em Z}}
{\lower.9pt\hbox{Z\kern-.4em Z}}{\lower1.2pt\hbox{Z\kern-.4em Z}}\else
{Z\kern-.4em Z}\fi}

 3

\twelvepoint
\baselineskip = 12pt
\rightline{UBCTP-94-005}
\rightline{June, 1994}
\vskip 0.5 truein
{\let\bf=\bigtenrm
\baselineskip=24pt
\centerline{\bf Equivariant Localization, Spin Systems and}
\centerline{\bf Topological Quantum Theory on Riemann Surfaces}}
\vskip 0.5truein
\centerline{{\bf Gordon W. Semenoff} and {\bf Richard J. Szabo}\footnote{*}
{\tenpoint This work was supported in part by the Natural
Sciences and Engineering Research Council of Canada.}}
\vskip 0.2truein
\centerline{\it Department of Physics}
\centerline{\it University of British Columbia}
\centerline{\it Vancouver, British Columbia, V6T 1Z1 Canada}
\vskip 2.0truein
\centerline{\bf Abstract}
\vskip 0.1truein

We study equivariant localization formulas for phase space path
integrals when the phase space is a multiply connected compact Riemann
surface. We consider the Hamiltonian systems to which the localization
formulas are applicable and show that the localized partition function
for such systems is a topological invariant which represents the
non-trivial homology classes of the phase space. We explicitly
construct the coherent states in the canonical quantum theory and show
that the Hilbert space is finite dimensional with the wavefunctions
carrying a projective representation of the discrete homology group of
the phase space. The corresponding coherent state path integral then describes
the quantum dynamics of a novel spin system given by the quantization of a
non-symmetric coadjoint Lie group orbit. We also briefly discuss the
geometric structure of these quantum systems.

\vfill\eject

\baselineskip=18pt

Recently there has been much discussion about the interrelationship between
equivariant cohomology and localization formulas [1--13].  For some quantum
systems, these formulas can be used to obtain an exact evaluation of
the quantum mechanical path integral $$Z(T)=\int_{T\Gamit}{\cal
D}x^\mu(t)~~{\rm det}^{1/2}\|\omega_{\mu\nu}\|
\e^{iS[x]}\eqno(1)$$
with action functional
$$S[x]=\int_0^Tdt~\left(\theta_\mu\dot{x}^\mu-{\cal
F}(H)\right)\eqno(2)$$ integrated over the space $T\Gamit$ of trajectories
on the phase space $\Gamit$ of a dynamical system. Here
$x^\mu$ are local coordinates on the $2n$ dimensional symplectic
manifold $\Gamit$ in which the symplectic structure is given locally
by $\omega={1\over2}\omega_{\mu\nu}(x)dx^\mu\wedge dx^\nu= d\theta$,
$\theta=\theta_\mu(x)dx^\mu$, and ${\cal F}(H)$ is a bounded
functional of some observable $H$ on $\Gamit$ which generates a global
symplectic circle action on the phase space
$$dH=-i_V(\omega)\eqno(3)$$ where $i_V$ is interior multiplication
which contracts differential forms with the globally defined Hamiltonian vector
field $V=V^\mu(x){\partial\over\partial x^\mu}$.  It has been argued
that, if the phase space admits, in addition to its symplectic
structure, a globally defined Riemannian structure which is invariant
under the U(1) action generated by $H$, then the path integral (1) can
be localized onto some effectively computable finite dimensional
expression, one example being the usual WKB formula [1--4].

These geometric localization techniques also provide conceptual geometric
approaches to quantum integrability [5], topological quantum field theory
[2,3] and Poincar\'e supersymmetric quantum field theory [6,7], and in the
appropriate instances the localization formulas give path integral
representations of index theorems [6], equivariant characteristic classes
[4], the infinitesimal Lefschetz number of Dirac operators [5], and the
Weyl and Kirillov characters of semi-simple Lie groups [2--4,8]. However,
most of the explicit examples where abelian equivariant
localization formulas have been shown to work deal with 2 dimensional
symplectic manifolds which are simply connected [2--4,8]. In these
cases the partition function (1) gives path integral representations
of the Weyl characters of the appropriate isometry groups acting on
these phase spaces [8]. Moreover, the Hamiltonians which obey the
equivariant localization constraints on these spaces are just families
of displaced harmonic oscillator Hamiltonians modified by the general
geometry of the phase space, and these are essentially the only
Hamiltonians to which these geometric constraints apply. Amongst other
things, in these cases one obtains a representation of how the phase
space geometry is realized explicitly in the underlying Hamiltonian
system and a probe into the geometrical structure of quantum systems.

In the following we shall study the equivariant localization formulas
on some 2 dimensional multiply connected symplectic manifolds.  We
consider the case where the phase space of a quantum system is a
compact Riemann surface $\Sigma_h$ of genus $h\geq1$, which has
non-trivial first homology group
$H_1(\Sigma_h;\IZ)=\bigoplus_{i=1}^{2h}\IZ$.  Since these spaces are
non-symmetric and so cannot be considered as K\"ahler manifolds
associated with coadjoint orbits of semi-simple Lie groups, as was the
case in previous examples, we expect the localizable systems in
these cases to be quite different than those studied previously. We
shall see that the Hamiltonian systems which satisfy the equivariant
localization criteria are rather small in number and define intriguing,
new types of quantum spin systems given by the quantization of the coadjoint
orbits $\prod_{i=1}^{2h}{\rm U}(1)$. The character
formula given by the partition function (1) in these cases gives path
integral representations of the homology classes of $\Sigma_h$. We
also examine the canonical quantum theory in the Schr\"odinger
representation and show that the Hilbert space of physical states is
finite dimensional. There we explicitly construct the coherent state
wavefunctions and show that they also carry a non-trivial
representation of the homology group of the Riemann surface. The coherent
state path integral defined by the propagator in these coherent states is
shown to give the exact result determined by (1). These
results provide examples of quantum systems on topologically
non-trivial phase spaces which are exactly solvable in both the path
integral and canonical quantization formalisms. We also examine
the localization onto equivariant characteristic classes and hence discuss
the geometrical structure of the quantum theory.

The localization formulas are reminescent of the Duistermaat-Heckman
theorem [9] which states that, if the function $H$ is as above and the
phase space is compact, the {\it classical} (finite-dimensional)
partition function is given exactly by the saddle-point approximation
$$\int_\Gamit{\omega^{\wedge n}\over n!}\e^{-\beta
H}={1\over\beta^n}\sum_ {x\in I(H)}{{\rm
det}^{1/2}_{(\mu\nu)}[\omega_{\mu\nu}(x)]\over{\rm det}^{1/2}
_{(\mu\nu)}\left[{\partial^2H\over\partial x^\mu\partial
x^\nu}\right]}
\e^{-\beta H(x)}\eqno(4)$$
where $I(H)$ is the set of critical points of $H$ which here are
assumed to be isolated. The localization formula (4) can be derived
using equivariant cohomological arguments [3,10]. Furthermore, when
(4) holds the Morse index of every critical point of $H$ must be
even, due to the circle action [11]. The standard example of the
localization formula (4) is when the phase space is the Riemann sphere
$S^2$ and $H$ is the height function on $S^2$ [3]. This is consistent
with the fact that in this case $H$ is a perfect Morse function with
even Morse indices. Since the same is true for the infinite
dimensional versions of the formula (4) [2--4,8], we begin our
analysis of the topologically non-trivial cases by considering the
effect of a multiply connected phase space on localization formulas
at the classical level.

To start, we review the situation for the case where the phase
space is the torus $\Sigma_1=S^1\times S^1$ viewed in 3-space as a
doughnut standing on the $xy$-plane and centered about the $z$-axis
[3]. Let $\tau\in{\cal H}$ be the modular parameter of the torus,
where ${\cal H}=\{z\in\IC:\Im z>0\}$ is the upper complex half-plane.
If $x^\mu$ are the angle coordinates on $S^1\times S^1$, then the
height function on $\Sigma_1$ can be written as
$$h_{\Sigma_1}(x^1,x^2)=r_2-(r_1+\Im(\tau)\cos x^1)\cos x^2\eqno(5)$$
where $r_1=|\Re\tau|+\Im\tau$ and $r_2=|\Re\tau|+2\Im\tau$. The
function (5) has 4 non-degenerate critical points on $\Sigma_1$: a
maximum at $(x^1,x^2)=(0,\pi)$, a minimum at $(0,0)$, and 2 saddle
points at $(\pi,0)$ and $(\pi,\pi)$ given by the bottom and top,
respectively, of the inner circle of the torus. If we take the
symplectic 2-form on $\Sigma_1$ to be the Darboux form
$\omega_D=dx^1\wedge dx^2$ and try to apply the Duistermaat-Heckman
theorem to this Hamiltonian system, then the formula (4) would give
$$\int_{\Sigma_1}\e^{-\beta
h_{\Sigma_1}}={1\over\beta\sqrt{\Im\tau}}\left[
r_2^{-1/2}\left(1+\e^{-2\beta
r_2}\right)+|\Re\tau|^{-1/2}\e^{-2\beta\Im\tau}
\left(1-\e^{-2\beta|\Re\tau|}\right)\right]\eqno(6)$$
However, a direct comparison of the first few terms of the power series
expansion in $\beta$ of the integral $\int_0^{2\pi}\!\!\int_0^{2\pi}dx^1~dx^2~
\e^{-\beta[r_2-\cos x^2(r_1+\Im(\tau)\cos x^1)]}$ with those of the
right hand side of (6) contradict this equality. This can be explained
by examining the Hessian matrix of the function (6) at each critical
point, which shows that while the Morse indices of the maximum and
minimum are 2 and 0, respectively, those of the saddle points are 1.
The observable (5) therefore generates no circle action on the torus
$\Sigma_1$.

This argument can be extended to the case where the phase space is a
hyperbolic Riemann surface $\Sigma_h=
\Sigma_1\#\cdots\#\Sigma_1$ ($h>1$), the $h$-fold connected sum of 2-tori.
$\Sigma_h$ can be viewed in 3-space as $h$ doughnuts stuck together
standing on end on the $xy$-plane and centered along the $z$-axis.
The height function on $\Sigma_h$ now has $2h+2$ critical points
consisting of $2h$ saddle points, 1 maximum and 1 minimum. Using the
results above for the torus one finds that the Duistermaat-Heckman
theorem (4) fails. Again this is because the $2h$ saddle points all
have Morse index 1 and so this height function doesn't generate any
U(1) action on $\Sigma_h$. Thus the height function restricted to a
compact Riemann surface can be used for Duistermaat-Heckman
localization only in genus $h=0$, and we see that the introduction of
more complicated topologies on 2 dimensional phase spaces restricts
even further the class of Hamiltonian systems to which classical
localization formulas can be applied.  In what follows we shall see
that this is also the case at the quantum level.

The path integral generalizations of the integration
formula (4) can be formally obtained using trajectory space equivariant
cohomology (with respect to the lifted circle action) and a supersymmetry
of the underlying Hamiltonian system [1--4,12]. The localization constraints
require that the phase space admit a globally defined metric tensor
$g={1\over2}g_{\mu\nu}(x)dx^\mu\otimes dx^\nu$ for which the
Hamiltonian vector field $V$ is a Killing vector $${\cal
L}_Vg={1\over2}\left(g_{\mu\lambda}\partial_\nu
V^\lambda+g_{\nu\lambda}
\partial_\mu V^\lambda+V^\lambda\partial_\lambda g_{\mu\nu}\right)dx^\mu\otimes
dx^\nu=0\eqno(7)$$ where ${\cal L}_V=di_V+i_Vd$ is the Lie derivative
along $V$. This phase space metric can then be lifted to give a
trajectory space metric tensor which defines dual 1-forms of vector
fields $W$ on the space of trajectories.  Standard equivariant
cohomological and supersymmetry arguments can then be used to formally
localize the path integral (1) onto the zeroes of $W$, and different choices
for $W$ will result in different localization schemes [8].

One choice for $W$ is $W^\mu=\dot{x}^\mu(t)-V^\mu(x(t))$, the
trajectory space Hamiltonian vector field corresponding to the action
functional (2).  This can be done provided that the critical points of
the action functional (2) are isolated and the Hamiltonian itself
generates a U(1) action on $\Gamit$ (i.e. ${\cal F}(H)=H$). The formal result
of evaluating the canonical localization integral in this case is the
well-known WKB localization formula [1--3] $$Z(T)=\sum_{x\in
I(S)}{{\rm det}^{1/2}\|\omega_{\mu\nu}\|\over{\rm det}
^{1/2}\|\delta^\mu_{~\nu}\partial_t-\partial_\nu(\omega^{\mu\lambda}
\partial_\lambda H)\|}\e^{iS[x]}\eqno(8)$$
where $\omega^{\mu\nu}$ is the matrix inverse of $\omega_{\mu\nu}$.
The formula (8) is the formal infinite dimensional generalization of
the Duistermaat-Heckman formula (4), and so under the geometric
criteria above we can always obtain a localization of (1) onto the
classical trajectories of the system.

A more general result, which does not require non-degeneracy of the
classical trajectories, can be obtained by setting
$W^\mu={1\over2}\dot{x}^\mu(t)$, which formally localizes the partition
function onto the time-independent modes of the phase space
trajectories. The final result is actually a localization onto
equivariant characteristic classes [4] $$Z(T)=\int
d\phi~\e^{-iTF(\phi)}\int_{\Gamit}{\rm ch}\left({T\over2} (\phi
H-\omega)\right)\wedge\hat{A}\left({T\over2}(\phi\Omega+R)
\right)\eqno(9)$$
where ${\rm ch}({T\over2}(\phi H-\omega))$ is the equivariant Chern
character of the 2-form $\omega$, $\hat{A}({T\over2}(\phi\Omega+R))$
is the equivariant $\hat{A}$-genus of the Riemannian manifold
$(\Gamit,g)$ and $F$ is the functional Fourier transform of the
Hamiltonian ${\cal F}(H)$.  The localization formula (9) is formally
an equivariant generalization of the Atiyah-Singer index of a Dirac
operator, in the equivariant cohomology associated with the
Hamiltonian vector field $V$ on $\Gamit$, and it reduces the infinite
dimensional expression (1) to the evaluation of some finite
dimensional integrals. Notice that, unlike the
WKB formula (8), the formula (9) is explicitly metric-dependent and so
some care must be taken in choosing the metric on $\Gamit$ to ensure
that the final result is independent of the phase space geometry, as
it should be\footnote{$^\dagger$}{\tenpoint\baselineskip=10pt See [8] and
[13] for some examples of the ambiguities associated with this
explicit metric dependence and their resolutions.}.

Notice that the localization formulas (8) and (9) are both derived from the
same geometric condition (7). We therefore expect, with the appropriate
restrictions, that these formulas are related to each other. Indeed, if
${\cal F}(H)=H$ and $S$ have only isolated critical points and each classical
trajectory can be contracted to a critical point of $H$ through a family of
classical trajectories, then we can take $W^\mu=V^\mu(x(t))$ to be the
lifted Hamiltonian vector field and formally obtain the localization
formula [3]
$$Z(T)=\sum_{x\in I_0(S)}{{\rm det}^{1/2}\|\omega_{\mu\nu}\|\over
{\rm det}^{1/2}\|\partial_\nu(\omega^{\mu\lambda}\partial_\lambda H)\|~
{\rm det}^{1/2}_\perp\|\delta^\mu_{~\nu}\partial_t-\partial_\nu(\omega^{\mu
\lambda}\partial_\lambda)\|}\e^{iTH(x)}\eqno(10)$$
where $I_0(S)$ is the set of time independent classical paths and $\det_\perp$
denotes the determinant with zero modes removed. The formula (10) can also
be derived from the WKB formula (8) using the Weinstein action invariant
[3], and from (9) by applying the ordinary Duistermaat-Heckman theorem (4)
to the phase space integral (9) [4]. In particular, if the localization
formula (10) holds, then as before the Hamiltonian $H$ admits only even
Morse indices [11].

Now let the phase space of a quantum system be the torus $\Sigma_1$
with modular parameter $\tau\in{\cal H}$, and consider a general
Hamiltonian system on $\Sigma_1$ to which the equivariant localization
constraints apply. We assume herein that all metrics have Euclidean
signature, and we shall now show that the geometric requirement (7) severely
limits the possible Hamiltonian functions on $\Sigma_1$.
$\Sigma_1$ can be regarded as the quotient $\IC/L_\tau$ of
its universal covering space $\IC$ by the free bi-holomorphic action
of the lattice group $L_\tau=\IZ\oplus\tau\IZ$ on $\IC$ [14]. Since
$\IC$ is simply connected it follows from Riemann uniformization that
the most general metric on it can be written globally in isothermal
form in terms of some conformal factor and the usual flat Euclidean
metric of the plane (after a possible diffeomorphism and Weyl
transformation of the coordinates).  The covering projection then
induces a metric on $\Sigma_1$, and we see therefore that the most
general metric on the torus can be written in terms of a flat K\"ahler
metric as $$g_\tau={\e^\varphi\over\Im\tau}dz\otimes d\zbar\eqno(11)$$
where the conformal factor $\varphi$ is a globally defined real-valued
function on $\Sigma_1$, and the complex structure on $\Sigma_1$ is
defined by the complex coordinate $z=x^1+\tau x^2$.

The normalization in (11) is chosen so that the volume of the torus
$${\rm vol}_{g_\tau}(\Sigma_1)=\int_{\Sigma_1}d^2x~\e^{\varphi(x)}=
(2\pi)^2v\eqno(12)$$ is finite and independent of the complex
structure of $\Sigma_1$. The metric (11) is further constrained by its
curvature
$$R(g_\tau)=\Im(\tau)\e^{-\varphi}\Delta^{(\tau)}\varphi\eqno(13)$$
which obeys the Gauss-Bonnet theorem
$$\int_{\Sigma_1}d^2x~\Delta^{(\tau)}\varphi(x)=0\eqno(14)$$ Here
$\Delta^{(\tau)}=\partial_z\overline{\partial}_z$ is the scalar
Laplacian
$$\Delta^{(\tau)}=\partial_{x^1}^2+|\tau|^{-2}\partial_{x^2}^2+2\Re(\tau)
|\tau|^{-2}\partial_{x^1}\partial_{x^2}\eqno(15)$$

The condition that the observable $H$ generates a globally integrable
isometry of (11) on $\Sigma_1$ implies that the Killing vectors
$V^\mu(x)$ must be single-valued functions under windings around the
non-trivial homology cycles of $\Sigma_1$. This means that these
functions must admit convergent 2-dimensional Fourier series
expansions $$V^\mu(x)=\sum_{n_\nu\in\IZ}V_{n_1,n_2}^\mu\e^{in_\nu
x^\nu}\eqno(16)$$ As we shall now demonstrate, these topological
restrictions severely limit the possible Hamiltonian systems to which
the equivariant localization constraints apply.

{}From (7) we see that the Killing equations for the metric (11) are
$$2\partial_{x^1}V^1+2\Re(\tau)\partial_{x^1}V^2+V^\mu\partial_
{x^\mu}\varphi=0$$
$$2\Re(\tau)\partial_{x^2}V^1+2|\tau|^2\partial_{x^2}V^2
+|\tau|^2V^\mu\partial_{x^\mu}\varphi=0$$
$$\partial_{x^2}V^1+\Re(\tau)(\partial_{x^2}V^2+\partial_{x^1}V^1)+|\tau|^2
\partial_{x^1}V^2+\Re(\tau)V^\mu\partial_{x^\mu}\varphi=0\eqno(17)$$
Substituting in the Fourier series (16) we find that (17) generates 2
coupled equations for the Fourier components of the Hamiltonian vector
field
$$(|\tau|^2n-\Re(\tau)m)V_{n,m}^1=|\tau|^2(m-\Re(\tau)n)V_{n,m}^2$$
$$(m-\Re(\tau)n)V_{n,m}^1=\left[(\Re(\tau)^2-\Im(\tau)^2)n-\Re(\tau)m\right]
V_{n,m}^2\eqno(18)$$ which hold for all integers $m$ and $n$. It is
easy to see from the coupled equations (18) that for $\tau\in{\cal
H}$, $V^1_{n,m}=V^2_{n,m}=0$ unless $n=m=0$. Thus the only
non-vanishing components of the Fourier expansions (16) are the
constant modes and the only Killing vectors of the metric (11) are the
generators of translations along the 2 independent homology cycles of
$\Sigma_1$. Notice that this result is completely independent of the
conformal factor in (11), and in fact could have been anticipated from
the onset: Although the torus inherits {\it locally} 3 isometries from
the plane (local rotations and translations), only the 2 translations
on $\Sigma_1$ are {\it global} isometries.  The independence of the
conformal factor is not surprising, since given any metric on a
compact space we can always form a U(1)-invariant metric by averaging
it over the circle group. However, the above derivation gives an
explicit geometric view of how the non-trivial topology of $\Sigma_1$
restricts the allowed U(1) actions on the phase space.

The symplectic structure, like the Riemannian structure, is invariant
under the generated U(1) action on $\Sigma_1$ (see (3)), so that
$${\cal
L}_V\omega=\partial_\mu\left(V^\lambda\omega_{\nu\lambda}\right)
dx^\mu\wedge dx^\nu=0\eqno(19)$$ For the Killing vectors above, (19)
implies that $\omega$ must be proportional to the Darboux 2-form
$\omega_D$, and so we take $$\omega_{\Sigma_1}=vdx^1\wedge
dx^2\eqno(20)$$ to be an associated volume form on $\Sigma_1$ for the
present geometry (see (12)). The Hamiltonian equations (3) then imply
that the observable $H$ is given by displacements along the homology
cycles of $\Sigma_1$ $$H_{\Sigma_1}(x^1,x^2)=h_1x^1+h_2x^2\eqno(21)$$
where $h_\mu$ are real-valued constants.

Besides defining a rather peculiar quantum system, (21) shows that the
allotted Hamiltonians determined from the geometric localization
constraints are in effect {\it independent} of the explicit form of
the phase space geometry and depend only on the topological properties
of the manifold $\Sigma_1$, i.e. (21) is independent of the conformal
factor $\varphi$ appearing in (11). This is completely opposite to
what occurs in the case of a simply connected phase space, where the
conformal factor of the given metric enters explicitly into the final
expression for the observable $H$ and the equivariant Hamiltonian
systems so obtained depend on the phase space geometry in a
non-trivial way [8,13]. In the present case the action functional (2)
with the Hamiltonian (21) obtained naturally from equivariant
localization defines a topological quantum theory on the torus, and
the corresponding partition function (1) will be a topological
invariant of the manifold $\Sigma_1$.

To explore some of the features of this topological quantum theory, we
set ${\cal F}(H)=H$ for the time being and consider the path integral
(1) on $\Sigma_1$ with symplectic 2-form (20) and Hamiltonian (21).
Although in the classical theory the Hamiltonian can be a multi-valued
function on $\Sigma_1$, to obtain a well-defined quantum theory we require
single-valuedness, under windings around the homology cycles of
$\Sigma_1$, of the time evolution generator
$\e^{-iH_{\Sigma_1}T}$ which defines the propagator (1).  This implies
that the constants $h_\mu$ must be quantized, $h_\mu\in h\IZ$ for some
$h\in\IR$, and then time propagation in this quantum system must be
defined in discretized intervals of the base time $h^{-1}$,
$T=N_Th^{-1}$ where $N_T\in\IZ^+$.

In the quantum theory the Hamiltonian (21) therefore represents the
winding numbers around the homology cycles of the torus, and the
partition function (1) is
$$Z^v_{\Sigma_1}(k,\ell;N_T)=\int_{T\Sigma_1}{\cal
D}x^\mu(t)~\exp\left[i\int
_0^{N_Th^{-1}}dt~\left(vx^2\dot{x}^1+h(kx^1+\ell
x^2)\right)\right]\eqno(22)$$ where $k$ and $\ell$ are integers. This
path integral can be evaluated directly and it gives
$$Z^v_{\Sigma_1}(k,\ell;N_T)=\e^{-ik\ell N_T^2/2v}\eqno(23)$$ Thus the
partition function of this quantum system represents the non-trivial
homology classes of the torus, through the winding numbers $k$ and
$\ell$ and the time evolution integer $N_T$. In fact, (23) defines a
family of 1-dimensional unitary representations of the homology group
of $\Sigma_1$ through the family of homomorphisms $Z_{\Sigma_1}^v(
\cdot,\cdot;N_T):H_1(\Sigma_1;\IZ)\to{\rm U}(1)\otimes{\rm U}(1)$
from the additive first homology group $\IZ\oplus\IZ$ of the torus
into a multiplicative circle group. Notice that the sum over all
winding numbers of the partition function (23) vanishes, so that the
associated homologically-invariant quantum theory is trivial.

In this case the WKB formula (8) directly gives the exact result (23),
while the localization formula (9) depends explicitly on the metric
(11) (i.e. on $\varphi$). It is here that the geometry of the phase
space enters explicitly into the quantum theory, if we demand that the
metric (11) be chosen so that the localization formula (9) coincides
with the exact result (23), as it should. In the case at hand (9) becomes
$$\eqalign{Z(T)&=\int_{\Sigma_1}{\rm ch}\left({T\over2}(H-\omega)\right)
\wedge\hat{A}\left({T\over2}(\Omega+R)\right)\cr&=\int_{\Sigma_1}d^2x~\int
d\psi^\mu~\e^{-iTH(x)+iT\omega_{\mu\nu}\psi^\mu\psi^\nu/2}
{}~{\rm
det}^{1/2}_{(\mu\nu)}\left[{{1\over2}(2\nabla_\mu^{g_\tau}V_\nu+R_{\mu\nu
\lambda\rho}(g_\tau)\psi^\lambda\psi^\rho)\over\sinh\left({T\over2}(2\nabla_
\mu^{g_\tau}V_\nu+R_{\mu\nu\lambda\rho}(g_\tau)\psi^\lambda
\psi^\rho)\right)}\right]\cr}\eqno(24)$$
where $\psi^\mu$ are anticommuting Grassmann variables. We substitute
into (24) the various expressions (11), (13), (20), (21) and the
Killing vectors from above, and then carry out the Grassmann
integrals. Comparing the result of this with the exact expression (23)
for the partition function, we arrive at a condition on the conformal
factor of the metric (11) $$\int_{\Sigma_1}d^2x~\e^{-iN_T(kx^1+\ell
x^2)}\left[1-{N_T^2(\ell\partial
_{x^1}\varphi-k\partial_{x^2}\varphi)^2\over4v^2\sinh^2\left({N_T\over2v}(\ell
\partial_{x^1}\varphi-k\partial_{x^2}\varphi)\right)}\right]^{1/2}=-{2i\over
N_Tv}\e^{-ik\ell N_T^2/2v}\eqno(25)$$
The Fourier series constraint (25) on
the metric is rather complicated and it represents the metric
ambiguity that we mentioned earlier. Notice, however, that (24) is
independent of the complex structure $\tau$.

The condition (25) can be used to check if a given phase space metric
really does result in the correct quantum theory (23), and this
procedure then tells us what quantum geometries are applicable to the
study of equivariant systems on the torus. For example, suppose we
tried to quantize a flat torus using equivariant localization. Then
from (13) the conformal factor would have to solve Laplace's equation
$\Delta^{(\tau)}\varphi=0$. Recall that $\varphi$ is assumed to be a
globally defined function on $\Sigma_1$, and so it must admit a
Fourier series expansion over $\Sigma_1$ as in (16). From (15) and the
Fourier series for $\varphi$ we see that Laplace's equation implies
that all Fourier modes of $\varphi$ except the constant modes vanish,
and so the left-hand side of (25) is zero. Thus a flat torus cannot be
used to localize the quantum mechanical path integral (22) onto the
equivariant Atiyah-Singer index (24). This simple example shows that
the condition (25), along with the Riemannian restrictions (12) and
(14), give a very strong probe of the quantum geometry of the torus.
Moreover, when (25) does hold, we can represent the equivariant
characteristic classes (24) in terms of the homomorphism (23) of the first
homology group of $\Sigma_1$.

We now examine the structure of the Hilbert space of this peculiar
quantum system. From (20) and its associated Poisson structure it
follows that the operators $z=x^1+\tau x^2$ and $\zbar=x^1+\overline
{\tau}x^2$ in the quantum theory obey the non-vanishing canonical
commutation relation $$[z,\zbar]={2\Im\tau\over v}\eqno(26)$$ In the
Schr\"odinger picture, (26) implies that the quantum states are
holomorphic functions $\Psi(z)$ and the generators of large
U(1) transformations around the homology cycles of $\Sigma_1$
are the unitary quantum operators
$$U(n,m)=\exp\left(2\pi(n+m\tau){\partial\over\partial z}+{\pi v\over
\Im\tau}(n+m\tbar)z\right)\qquad;\qquad n,m\in\IZ\eqno(27)$$
The operators (27) generate simultaneously the winding transformations
$z\to z+2\pi(n+m\tau)$ and $\zbar\to\zbar+2\pi(n+m\tbar)$ and, by the
above arguments, the quantum states should be invariant under their
action on the Hilbert space. Solving this invariance condition will
then give a representation of the equivariant localization constraints
directly in the Hilbert space.

In general, products of the operators (27) do not commute and differ from
their reversed action by a ${\rm U}(1)\times{\rm U}(1)$ 2-cocycle.
Using the Baker-Campbell-Hausdorff formula, we find that they obey the
clock algebra
$$U(n_1,m_1)U(n_2,m_2)=\e^{2\pi iv(n_2m_1-n_1m_2)}U(n_2,m_2)U(n_1,m_1)
\eqno(28)$$
and that their action on the quantum states of the theory is
$$U(n,m)\Psi(z)=\exp\left[{\pi
v\over\Im\tau}\left(\pi|n+m\tau|^2+(n+m\tbar)z
\right)\right]\Psi(z+2\pi(n+m\tau))\eqno(29)$$
If the parameter $v=~{\rm vol}_{g_\tau}(\Sigma_1)/(2\pi)^2$ is an
irrational number, then it follows from the clock algebra (28) that
the U(1) generators act as infinite dimensional raising operators in
(29) and so the Hilbert space of quantum states in this case is
infinite dimensional.

More interesting, however, is the case where the volume of the torus is
quantized so that $v={v_1\over v_2}$, $v_1,v_2\in\IZ^+$, is
rational-valued.  In this case the cocycle relation (28) shows that
$U(v_2n,v_2m)$ commutes with all of the other U(1) generators and the
time evolution operator, and so they can be simultaneously
diagonalized over the same basis of states:
$U(v_2n,v_2m)\Psi(z)=\e^{i\eta_{n,m}}\Psi(z)$, where the phases are
${\rm U}(1)\times{\rm U}(1)$ 1-cocyles $\eta_{n,m}\in
S^1$.  This expresses explicitly the invariance of the quantum states
under the U(1) action on the phase space, and using (29) it can be
written as $$\Psi(z+2\pi(n+m\tau))=\exp\left[i\eta_{n,m}-{\pi
v_1\over\Im\tau}\left(
\pi v_2|n+m\tau|^2+(n+m\tbar)z\right)\right]\Psi(z)\eqno(30)$$
The quasi-periodicity condition (30) can be uniquely solved in terms
of the Jacobi theta functions [14] $$\Theta^{(D)}\pmatrix{c\cr
d\cr}(z|\tilde{\Omega})=\sum_{
\{n^\ell\}\in\IZ^D}\exp\left[i\pi(n^\ell+c^\ell)\tilde{\Omega}_{\ell
p}(n^p+c^p)+2\pi i(n^\ell+c^\ell)(z_\ell+d_\ell)\right]\eqno(31)$$
where $c^\ell,d_\ell\in[0,1]$. The functions (31) are holomorphic
functions of $\{z_\ell\}\in\IC^D$ for
$\tilde{\Omega}=[\tilde{\Omega}_{\ell p}]$ in the Siegal upper half-plane.

{}From the doubly semi-periodic behaviour of (31) [14] we find that the
constraint (30) is solved by the $v_1v_2$ independent holomorphic
functions $$\Psi_{p,r}\pmatrix{c\cr
d\cr}(z)=\e^{-(v/4\Im\tau)z^2}\Theta^{(1)}
\pmatrix{{c+2\pi v_1p+v_2r\over2\pi v_1v_2}\cr d\cr}(v_1z|2\pi
v_1v_2\tau)\eqno(32)$$ where $p=1,2,\ldots,v_2$ and
$r=1,2,\ldots,v_1$. The winding transformations (29) can then be
represented by finite dimensional matrices
$$U(n,m)\Psi_{p,r}\pmatrix{c\cr d\cr}(z)=\sum_{p'=1}^{v_2}\left[
U(n,m)\right]_{pp'}\Psi_{p',r}\pmatrix{c\cr d\cr}(z)\eqno(33)$$ with
$[U(n,m)]_{pp'}=\e^{2\pi i(cn-dm+\pi v_1nm)/v_2}\delta_
{p+m,p'}$. The Hilbert space is now $v_1v_2$-dimensional and the
quantum states carry a $v_2$-dimensional projective representation of
the clock algebra (29) (i.e. of the equivariant localization
constraints (3) and (7)) with 2-cocycle $\alpha_2={v_1\over
v_2}(n_2m_1-n_1m_2)$. This shows how the U(1) equivariant localization
constraints and the toroidal restrictions are realized in the
canonical quantum theory, since these imply that the only possible
unitary operators on the Hilbert space are combinations of the
generators (27).

Notice that although the 2 parameters $c$ and $d$ in the wavefunctions
(32) appear as free variables, one of them can be eliminated by
requiring that the Hamiltonian (21) in this basis of states does
indeed lead to the correct propagator (23). (23) should be equal to
the trace of the time evolution operator in the finite dimensional
vector space spanned by the wavefunctions (32) $${\rm
Tr}~\e^{-iTH_{\Sigma_1}(k,\ell)}=\sum_{p=1}^{v_2}\sum_{r=1}^{v_1}
\left(\Psi_{p,r},\e^{-iTH_{\Sigma_1}(k,\ell)}\Psi_{p,r}\right)\eqno(34)$$
where the inner product on the subspace spanned by the coherent states
(32) is defined by the usual coherent state measure [15]
$$(\Psi_1,\Psi_2)=\int_{\Sigma_1}d^2z~\e^{(v/2\Im\tau)z\zbar}\Psi_1^*(\zbar)
\Psi_2(z)\eqno(35)$$
With the measure (35) we find that the vectors (32) are orthonormal.
Substituting $\e^{-iTH_{\Sigma_1}(k,\ell)}=[U(\ell,-k)]^{N_Tv_2/2\pi
v_1}$ into (34), using (33) and (35), and then comparing this final
result for (34) with the exact one (23), we find that the parameter
$d$ in the wavefunctions (32) can be determined by $d=(k\ell
N_T-2ck)/2\ell$. The remaining degree of freedom $c$ can then be fixed
by requiring that the wavefunctions (32) be modular invariants [14],
as they should be since the topological quantum theory defined by (22)
is independent of the phase space complex structure.

With these parameter values, the propagator (34) then corresponds to the
canonical coherent state path integral [15]
$$\eqalign{Z_{\Sigma_1}^v(k,\ell;N_T)=\int_{T\Sigma_1}&\left(\prod_tdz(t)~
d\zbar(t)~\e^{(v/2\Im\tau)z(t)\zbar(t)}\right)\cr&\times\exp\left[{1\over
2i\Im\tau}\int_0^{N_Th^{-1}}dt~\left\{{v_1\over2v_2}\left(\zbar\dot{z}
-z\dot{\zbar}\right)+ih\left((\ell-\tau k)\zbar-(\ell-\tbar k)z\right)
\right\}\right]\cr}\eqno(36)$$
The coherent state path integral (36) gives the quantization of the unusual
spin system defined by (21) in terms of the quantized coadjoint orbit
${\rm U}(1)\times{\rm U}(1)=S^1\times S^1$. The points on this orbit are in
one-to-one correspondence with the coherent state representation of the
projective clock algebra (28) of the discrete homology group of the torus.

The possibilities of using arbitrary functionals ${\cal F}
(H_{\Sigma_1})$ of the observable (21) are far more restrictive here
than in the case where the phase space is simply connected. There we
require generally only that $\cal F$ be bounded from below, or else
the propagator (1) may become ill-defined as a tempered distribution
[8], while in the case at hand we need in addition the requirement
that $\cal F$ is formally a periodic functional of the observable
(21), for the same reasons as before. In general this will impose no
quantization condition on the time translation $T$, as it did above.
These remarks show, for example, that one cannot equivariantly
quantize a free particle (with a compactified momentum range) on the
torus. The same is true of the height function (5), as anticipated,
and in these cases the periodicity of the Hamiltonian results in a
much better defined propagator.

We conclude by discussing how the explicit analysis above generalizes to the
case where the phase space is a hyperbolic Riemann surface $\Sigma_h$. For
$h>1$, $\Sigma_h$ can be regarded as the quotient ${\cal H}/F_h$ of the
Poincar\'e upper half-plane $\cal H$, with a hyperbolic metric, by the free
bi-holomorphic action of a discrete Fuchsian group $F_h$ [14]. The metric
$g^{ }_{\Sigma_h}$ induced on $\Sigma_h$ by the universal bundle projection
then involves a conformal factor $\varphi$ as in (11) and a constant negative
curvature K\"ahler metric. Choosing a canonical homology basis $\{a_\ell,b_
\ell\}_{ \ell=1}^h\subset H_1(\Sigma_h;\IZ)$, the condition that the Killing
vectors $V^\mu(x)$ be globally defined on $\Sigma_h$ means that they
must be single-valued under windings around these homology cycles, or
equivalently that
$$\oint_{a_\ell}dV^\mu=\oint_{b_\ell}dV^\mu=0\eqno(37)$$
As before, $\Sigma_h$ inherits 3 local isometries from the Poincar\'e
upper half-plane. However, only the 2 quasi-translations on $\cal H$
become global isometries of $\Sigma_h$ and they can be expressed in
terms of the canonical homology basis. This global isometry condition,
from (37) and the Killing equation $di_V(g^{ }_{\Sigma_h})=-i_V(dg^{
}_{\Sigma_h})$, implies that the Hodge decomposition over $\Sigma_h$ of
the 1-form $i_V(g^{ }_{\Sigma_h})$ dual to the Hamiltonian vector field
$V$ is simply $$i_V(g^{
}_{\Sigma_h})=\sum_{\ell=1}^h\left(V_1^\ell\alpha_\ell+V_2^\ell
\beta_\ell\right)\eqno(38)$$
where $\{\alpha_\ell,\beta_\ell\}\subset H^1(\Sigma_h;\IZ)$ is an
orthonormal basis of harmonic 1-forms on $\Sigma_h$ which are Poincar\'e-dual
to the canonical homology basis. The Killing vectors dual to (38) generate
translations along the homology cycles of $\Sigma_h$.

The symplecticity condition (19) now becomes
$$di_V(\omega)=\sum_{\ell=1}^hd\left(\bar{\omega}*\left[V_1^\ell\alpha_\ell+
V_2^\ell\beta_\ell\right]\right)=0\eqno(39)$$ where $*$ denotes the
Hodge duality operator with respect to the metric $g^{ }_{\Sigma_h}$
of $\Sigma_h$ and $\omega=\bar{\omega}(x)dx^1\wedge dx^2$. (39)
implies that the function $\bar{\omega}(x)$ is constant on $\Sigma_h$,
and thus the solutions to the Hamiltonian equations (3) have the form
$$H_{\Sigma_h}(x)=\sum_{\ell=1}^h\int_{C_x}\left(h_1^\ell\alpha_\ell+
h_2^\ell\beta_\ell\right)\eqno(40)$$ where $h_\mu^\ell$ are
real-valued constants and $C_x\subset\Sigma_h$ is a simple curve from
some fixed basepoint to $x$. As before, single-valuedness of the time
translation generator requires that $h_\mu^\ell=n_\mu^\ell h$, for
some $n_\mu^\ell\in\IZ$ and $h\in\IR$, and the time translations are
again the discrete intervals $T=N_Th^{-1}$. Thus the Hamiltonian (40)
represents the windings around the non-trivial homology cycles of
$\Sigma_h$ and the action functional (2) with this Hamiltonian defines
a topological quantum theory.  The partition function will then again
represent the homology classes of $\Sigma_h$ through a family of
homomorphisms from $\bigoplus_{i=1}^{2h}\IZ$ into ${\rm
U}(1)^{\otimes2h}$.

The rest of the analysis performed above for the 2-torus can now be
carried through for the case at hand, except that now the
coordinatization of $\Sigma_h$ is far more intricate because its
complex structure involves $3h-3$ complex parameters, as opposed to
just 1. The conformal factor of $\Sigma_h$ obeys a volume constraint
similar to (12), and it will be further constrained again by the
Gauss-Bonnet theorem which now reads $\int_{\Sigma_h}\sqrt{g^{
}_{\Sigma_h}}d^2x~R(g^{ }_{\Sigma_h})=4\pi(1-h)$.  When the volume
parameter is quantized as before, the Hilbert space of physical states
will be $(v_1v_2)^{3h-3}$ dimensional and the coherent state
wavefunctions, which can be expressed in terms of $(3h-3)$-dimensional
Jacobi theta functions, will in addition carry a $(v_2)^{3h-3}$
dimensional projective representation of the discrete homology group
of $\Sigma_h$ (i.e. of the equivariant localization constraint
algebra).

We see that the general feature of abelian equivariant localization of
path integrals on multiply connected compact Riemann surfaces is that it
leads to a topological quantum theory whose associated topologically invariant
partition function represents the non-trivial homology classes of the phase
space. The states of the finite dimensional Hilbert space also carry a
multi-dimensional representation of the discrete homology group, and the
localizable Hamiltonian systems on these phase spaces are rather
restricted and unusual. The U(1)-invariant symplectic 2-forms that one
obtains in these cases are non-trivial elements of
$H^2(\Sigma_h;\IZ)=\IZ$, as in the simply connected cases [8], and it
is essentially the global topological features of these phase spaces
which results in these rather severe restrictions. The coherent state
quantization of these systems shows that the path integral is the
coadjoint orbit quantization of an unusual
spin system on the Riemann surface. Moreover, these spin systems are
exactly solvable and we have obtained explicit solutions of the
quantum theory both from the point of view of path integral
quantization on the space of trajectories and of canonical holomorphic
quantization of the phase space. Although in these examples and
those which arise on simply connected phase spaces the integrable
quantum systems are rather trivial, we expect that more non-trivial
examples will emerge when instead of circle actions one considers the
Poisson action of some nonabelian Lie group acting on the phase space.
Nonabelian generalizations of the equivariant localization formulas
have been discussed in [5] and [16], and in these versions the
structure of the quantum representations
discussed above and in [8] and [13] are expected to be much richer. On
these spaces one might then obtain intriguing path integral
representations of the groups involved.

\vfill\eject

\baselineskip=12pt

\centerline{\bf References}

\bigskip

\line{\hfill}
\item{[1]} M. Blau, E. Keski-Vakkuri and A. J. Niemi, Phys. Lett. {\bf
B246} (1990), 92.
\line{\hfill}
\line{\hfill}
\item{[2]} A. J. Niemi and P. Pasanen, Phys. Lett. {\bf B253} (1991), 349.
\line{\hfill}
\line{\hfill}
\item{[3]} E. Keski-Vakkuri, A. J. Niemi, G. W. Semenoff and O. Tirkkonen,
Phys. Rev. {\bf D44} (1991), 3899.
\line{\hfill}
\line{\hfill}
\item{[4]} A. J. Niemi and O. Tirkkonen, Ann. Phys. (1994), in press.
\line{\hfill}
\line{\hfill}
\item{[5]} A. J. Niemi and K. Palo, Mod. Phys. Lett. {\bf A8} (1993), 2311.
\line{\hfill}
\line{\hfill}
\item{[6]} A. Hietam\"aki, A. Yu. Morozov, A. J. Niemi and K. Palo, Phys.
Lett. {\bf B263} (1991), 417; A. Hietam\"aki and A. J. Niemi, Phys. Lett.
{\bf B288} (1992), 321.
\line{\hfill}
\line{\hfill}
\item{[7]} A. Yu. Morozov, A. J. Niemi and K. Palo, Nucl. Phys. {\bf B377}
(1992), 295; K. Palo, Phys. Lett. {\bf B321} (1994), 61.
\line{\hfill}
\line{\hfill}
\item{[8]} R. J. Szabo and G. W. Semenoff, Nucl. Phys. {\bf B421}, No. 2
(1994).
\line{\hfill}
\line{\hfill}
\item{[9]} J. J. Duistermaat and G. J. Heckman, Inv. Math. {\bf 69} (1982),
259.
\line{\hfill}
\line{\hfill}
\item{[10]} N. Berline and M. Vergne, Duke Math. J. {\bf 50} (1983), 539;
Am. J. Math. {\bf 107} (1985), 1159; M. F. Atiyah and R. Bott, Topology
{\bf 23} (1984), 1.
\line{\hfill}
\line{\hfill}
\item{[11]} F. Kirwan, Topology {\bf 26} (1987), 37.
\line{\hfill}
\line{\hfill}
\item{[12]} M. F. Atiyah, Asterisque {\bf 131} (1985), 43; J. M. Bismut,
Commun. Math. Phys. {\bf 98} (1985), 213; {\bf 103} (1986), 127.
\line{\hfill}
\line{\hfill}
\item{[13]} H. M. Dykstra, J. D. Lykken and E. J. Raiten, Phys. Lett.
{\bf B302} (1993), 223.
\line{\hfill}
\line{\hfill}
\item{[14]} M. Schlichenmaier, {\it An Introduction to Riemann Surfaces,
Algebraic Curves and Moduli Spaces}, Springer-Verlag (Berlin) (1989).
\line{\hfill}
\line{\hfill}
\item{[15]} A. M. Perelomov, {\it Generalized Coherent States and Their
Applications}, Springer-Verlag (Berlin) (1986).
\line{\hfill}
\line{\hfill}
\item{[16]} E. Witten, J. Geom. Phys. {\bf 9} (1992), 303; A. J. Niemi and
O. Tirkkonen: Equivariance, BRST and Superspace, University of Uppsala
preprint UU-ITP 01/94 (1994).

\end